# Decentralized social networks and the future of free speech online

Tao Huang

*City University of Hong Kong, Hong Kong, China*

A B S T R A C T

Decentralized social networks like Mastodon and BlueSky are trending topics that have drawn much attention and discussion in recent years. By devolving powers from the central node to the end users, decentralized social networks aim to cure existing pathologies on the centralized platforms and have been viewed by many as the future of the Internet. This article critically and systematically assesses the decentralization project's prospect for communications online. It uses normative theories of free speech to examine whether and how the decentralization design could facilitate users' freedom of expression online. The analysis shows that both promises and pitfalls exist, highlighting the importance of value-based design in this area. Two most salient issues for the design of the decentralized networks are: how to balance the decentralization ideal with constant needs of centralization on the network, and how to empower users to make them truly capable of exercising their control. The article then uses some design examples, such as the shared blocklist and the opt-in search function, to illustrate the value considerations underlying the design choices. Some tentative proposals for law and policy interventions are offered to better facilitate the design of the new network. Rather than providing clear answers, the article seeks to map the value implications of the design choices, highlight the stakes, and point directions for future research.

## 1. Introduction

Decentralization seems to be the theme of the next generation web.[1] Disappointed with the centralized commercial social media such as Facebook and Twitter, a growing number of people has resorted to alternative networks that are decentralized in structure. Twitter has witnessed a Diaspora of its users toward Mastodon. The release of BlueSky has gained traction in the social media world. Even the commercial giant Meta has launched a new social media platform, Threads, and announced it will connect with the decentralized networks.

The decentralized social network has been widely viewed as a cure to its centralized counterpart, which is owned by corporate monopolies, funded by surveillance capitalism, and moderated according to rules made by the few ([1], 2–3). The tremendous and unchecked power of those giant platforms was seen as a major threat to people's rights and freedoms online. The newly emergent decentralized social networks, through infrastructural redesign, create a power-sharing scheme with the end users, so that it is the users themselves, rather than a corporate body, that determine how the communities shall be governed. Such an approach has been hailed as a promising way of curbing the monopolies and empowering the users. It was expected to bring more freedom of speech to individuals, and the vision it underscores – openness rather than walled-gardens, bottom-up rather than top-down – represents the future of the Internet ([2], 115).[2]

The discussion of the decentralization project is trending, but it is too limited for two aspects. First, the current research focused primarily on the technical features, difficulties and improvements [3–5], lacking systematic and critical review on the project's normative implications ([6], 2). This is insufficient because without normative insights from the fields of policy, law, and ethics, it is unlikely that the new technology will achieve its expected goals. If decentralized networks have the potential to become the future of online communications, we need to know why this is so and how it can be achieved. All the utopian and dystopian analyses of the decentralized future are only possibilities: what matters is the choices we make in using and designing the technology [7]. Value commitments must be carefully examined and considered in the design process.

Second, for those studies that touched upon the normative implications of decentralize social networks, most has addressed the issues from economic, social or cultural perspectives. For instance, [8] addressed the

*E-mail address:* taohuang@cityu.edu.hk.

[1] It has been described as a constitutional moment, *see* [60], 4.

[2] The chief of Instagram said decentralization is the future of social networks, *see* [124].

decentralization's effect on community building and norm development; [9] discussed the decentralized structure's impact upon participatory culture of online communities; [10] examined challenges for governing the decentralized social media from socio-economic perspectives. Legal analyses, especially those evaluating from the fundamental legal right of free speech, is very few. The few works that explore decentralization's impact on free speech are parochial and piecemeal, in the sense that they have pointed out some of the challenges of decentralization while ignoring other aspects, and their examination is collections of ad hoc thoughts rather than systematic and consistent inquiries guided by theories. For example, [11,12] stressed the filter bubble risk, while [13] and [14] focused on the moderation challenges. Without guidance from normative theories, their works not only failed to grasp the full picture of decentralized networks' promises and perils, but also are unable to answer how the challenges they identified might threaten the value commitments we hold toward the new technology.

This article tries to fill part of the gap by incorporating the free speech scholarship into the current debate over the technical design of decentralized social networks. It is the first piece that uses normative theories of free speech to evaluate the decentralization project in a systematic and critical way. Social media, as Internet-based services that host, link, and organize users' content and interactions ([15], 18; [16], 747),[3] constitute the most vital space for human communications in the digital age. Originating from the web 2.0 applications that are more interactive and participatory than their predecessors ([16], 746; [17]), social media platforms enable individual expressions and social interactions, facilitate information dissemination and knowledge sharing, as well as promote cultural exchange and engagement. They have been described as "the modern public square" ([18], 1737). Due to their crucial role in public discourse, we hold normative commitment that social media facilitate freedom of speech online and this is one major incentive that drives the development of the decentralized networks. But we can't just expect that the decentralized structure itself is sufficient to realize the value commitment: think about that decentralization was once the mainstream structure of the early years Internet and then largely failed. We need to know, instead, what type of design of the decentralized networks can really bring the free speech future we are longing for.

The article uses three major free speech values – knowledge, democracy, and autonomy – to test and guide the design choices. It tries to locate the potential benefits and drawbacks of the decentralization structure to the free speech values, and to link the relevant value conflicts with currently debated design questions. Due to space limit, this article will not prescribe which specific design choices are preferable, but rather seeks to map the implications of the design choices, highlight the stakes, and point directions for future research.

Part 2 offers brief background for readers who are not familiar with the decentralized social network: it introduces its structural characteristics, typical types, and current states in the industry. Part 3 illustrates the normative ground for incorporating free speech theories into value-based design: why value-based design is necessary for technical development, and why I choose the three values (knowledge, democracy, and autonomy) to guide the design process. Part 4 outlines the promises and pitfalls of the decentralization structure for free speech values. It shows that all the three major values can be served and disserved by the decentralized design. This analysis calls for a sober and critical stance toward the overheated hail of decentralization, as well as underscores the importance for design choices – if decentralization itself is no guarantee of desirable outcomes, then it is the detailed design choices that dictate which online space we will get.

Based on the value analysis in Part 4, the next two parts highlighted two pressing concerns in the design of the decentralized networks. Part 5 argues that the decentralization design of social networks constantly faces the phantom of (re)centralization. The need of epistemic management, content moderation, and cross-community dialogue generates the centripetal force among the decentralized nodes on the networks. Part 6 warns that the autonomy and agency granted to users by decentralization may only become illusion of control or theatre of autonomy. Real autonomy requires far more than the mere increase of formal choices; that's the lesson we have learned from the consent/control dilemma of privacy law. Thus, how to balance the decentralization ideal with the centralization need, and how to truly empower the users in the governance are two ongoing themes in the design debate of the decentralized networks. Part 7 lists some examples of the design choices that reflect those themes. The examples include whether to detach moderation from data hosting, whether to provide default servers, how to coordinate moderation across servers, whether the functions of federation and search should be opt-in or opt out, and how to design the federated timeline of content.

After elaborating on the value concerns and design choices in the decentralized social networks, Part 8 offers some tentative suggestions of how law and policy could intervene, in order to make the design of the networks more in line with the values identified. These suggestions are neither comprehensive nor final, but subject to further revision and specification in future debate. The last part concludes the article.

## 2. Decentralized social networks: features and types

Technically speaking, the difference between decentralized networks and centralized networks lies primarily in their topological structures. Topology refers to how the nodes in a communication network are connected to each other, or, simply put, its logical layout ([8], 1192). Centralized networks are characterized by the leading role of a central node, which governs and controls all the other nodes (ends, or peripheries) in the network; the communications and data exchange between any two nodes must be mediated or hosted by the central node. All the mainstream/popular social media, such as Facebook, Twitter, and TikTok, follow this model. The central node, which is the giant platform owned by a corporate entity, stores the data of all end-users, hosts the communications among the users, and moderate users' speech according to uniform rules made by the central.

Decentralized networks, by contrast, are governed upon "relationships between individuals in flat social structures, enabled through technologies which support connection and coordination, without any central control" ([19], 3). On these networks, users/nodes form communities of their own without the intermediation of the central node/platform. In other words, either one or both of the two primary functions of social networks – data hosting and content moderation – will be devolved from central control to the users. Data hosting is the function of storing and maintaining users' data – this function lies in the hardware level and constitutes the final point of control for the networks. Content moderation is another key function of social networks, referring to the management of user-generated content – it is usually done in the application level and is more visible to users. Different types of decentralized social networks can be categorized according to the various ways the two functions have been devolved. At least three mainstream types can be summarized as follows:

(1) Devolve the power of content moderation but not the power of data hosting from the central. One representative proposal is the Middleware plan made by Fransis Fukuyama and his team: the proposal is to outsource content moderation service to third-party providers and let users freely choose from those providers ([20], 34; [21], 5). Under this proposal, however, the users' data is still hosted by the central platform. There are other similar proposals within this category, such as Wolfram

[3] Major social media platforms include, e.g., Facebook, Twitter (X), Instagram, Weibo, YouTube, Reddit, TikTok, Threads. For this article's purposes, I focus on social media that build and maintain social networks on the basis of user profiles, posts, and communications. Other platforms, then, such as those only provide instant messaging services (e.g., WhatsApp), are not the focus of this article's discussion.

[22], Article 19 (2021) and Stasi [23].

(2) Devolve both powers of data hosting and content moderation to users in a bundled way. That means, data hosting and content moderation will be open to users or any providers, but the two functions cannot be separated. Any individual or organization can choose to provide the two services in the bundled way, and users can freely choose which bundle of services they would like to use. Representative example of this kind is the Mastodon network, building upon the ActivityPub protocol (the networks built upon this protocol have been called Fediverse).

(3) Devolve both data hosting and content moderation to users in an unbundled way, so that the two functions can be performed by different entities or individuals. Under this model, a user can choose one data storage service to store his/her data while at the same time subscribe to another service provider who will moderate the content he/she can see on the network. BlueSky (and the protocol it has built upon, AT protocol) is a typical example that adopts this model.

These three models are currently dominating the discussion about decentralized social networks, especially the second and the third. The first, Middleware model receives less attention because it still relies on centralized platforms to store user data, thus leaving the concerns of data surveillance and single point of failure unaddressed.

Apart from those mainstream models, there is a radical approach of decentralization: the peer-to-peer (P2P) structure. This model devolves the powers of data hosting and content moderation into each and every end user on the network, making each user both a client and a server [3]. Communications occur in a peer-to-peer or end-to-end way. In contrast to the above three models which empower the users in a *collective* way, enabling users to choose from different providers of data hosting service and content moderation service, the P2P model requires each *individual* user to host data and moderate content. This is the most decentralized topology for social networks, and is the most resistant against censorship, data breach, and manipulation. However, asking each user to maintain and manage a server is utopian, at least in the short term, because the technical knowhow, economic cost, and time engagement required are unbearable for ordinary users [24]. Such high threshold renders the P2P model prevalent in small circles at best, rather than a major type of social network that can be widely adopted ([25], 227).

| | **The ways of devolving powers from the central to the users** | | | |
|---|---|---|---|---|
| | Devolving power to users in a collectively way | | | Devolving power to each individual user |
| | Devolving only power of content moderation | Devolving content moderation and data hosting in a bundle | Devolving content moderation and data hosting separately | |
| **Examples** | Middleware | Mastodon | BlueSky | P2P network |

It should be noted that some decentralized social networks have relied upon blockchain – itself a decentralization technology. Blockchain is a network technology that enables and validates transactions without human intervention and intermediary control: it is trustless and permissionless ([26], 3). Social media applications that use blockchain technology include, e.g., *Minds, Steemit*, and *Peepeth*. Several advantages can be gained from using blockchain to build social networks, such as "enhanced privacy by minimizing third-party access to personal data, improved resilience and availability by eliminating a single point of failure, and the promotion of collaborative, community-driven interactions that empower users to have greater control over their network experience" [27].

Blockchain can be incorporated into social media to achieve decentralization in varying degrees, and it has been used mainly in two aspects: data storage or monetization ([28]b). The decentralized storage of data on blockchain facilitates resistance to censorship, enhanced user control, and data immutability [29]. But the immutability of on-chain data makes the management of bad content more difficult ([30], 19), and it also makes the technology hard to scale because user data on social media is mutable and ephemeral ([28]b). Another function brought by blockchain is the monetization of user content or attention, through the use of cryptocurrencies. Such tokenized system may incentivize user contribution [29] and even has the potential of emerging as a new business model for social media [11]. As for now, the tokenization model is still in its infancy, and we still need time to observe whether it could become a sustainable model for the industry. As monetization mainly relates to the business model of social media, this article will focus on the impact of data immutability on free speech, especially the epistemic value of speech (see part 4.1).

## 3. Value-based design and free speech values

### 3.1. Why value-based design?

Designing the structure, functionality, and interface of the decentralized social networks is new enterprise. Such design should be guided by normative values because first, social networks are so important for our social and public life. As commentators state, "the largest online platforms, such as Facebook and Google, exercise more power over our right to free expression than any court, king, or president ever has—in view of the very significant percentage of human discourse that occurs within the boundaries of these [platforms]" ([31], 38). People rely on those online networks to make friends, receive news, and engage in discussions, and how they are structured would significantly influence our experience and interests online.

Second, social networks cannot fulfill their important public functions automatically. Technologies in our time have been continuously used to entrench the power imbalance between users and businesses in a structural and systematic way ([32], 176, 187). For technologies to empower, rather than deprive or manipulate, the vulnerable individuals, and for them to realize the expected functions, value propositions must be considered and incorporated into the design process ([33], 1).

That's why value-based design or value-sensitive design is indispensable in this field. Value-based design can be defined as "an approach to the design of technology that accounts for human values in a principled and comprehensive manner throughout the design process" ([34], 1125). This process should be participatory and inclusive because technicians cannot make decisions about what values to include, how to resolve value conflicts, and how to translate values into design instructions solely by themselves. Rather, input from philosophers, social scientists, lawyers, and ordinary users is needed.

This article contributes to the value-based design of decentralized social networks by incorporating free speech theories from constitutional law scholarship into the design considerations of the decentralization technology. The following section describes what values of free speech should be included and what norms they entail.

### 3.2. What free speech values?

Value-based designers and researchers should make explicit claims about what values have been considered and why; they should acknowledge the limits of their choice as well as possible approaches to deal with value conflicts ([34], 1127; [35], 2). First, I select values from the field of free speech because social media have been widely considered as important forums for exercising free speech rights ([36], para. 158–160). This is also one of the major rationales for supporting the decentralization project. Of course, this does not mean that other values, such as privacy and data protection, are not important and do not need to be incorporated into the design process. The scope of this article, however, does not allow for a comprehensive study of those values but rather leaves them to future research.

Second, value-based design can be generally adopted as a process of specification: that means, it is a mechanism from the abstract to the specific. In other words, values in the high level must first be translated into norms in the mid-level, and then further specified into design requirements that can be directly applied in the design process ([37], 259). The high-level values and the mid-level norms may be much less controversial than the low-level design requirements. That's because, when designers tried to translate normative commitments into practical design requirements, they must strike balance between conflicting values as well as make tradeoffs among values and practical needs or limitations. That explains the reason why the design of decentralized networks has to make key choices as to which value to prioritize, what does a value mandate in practice, and how to achieve the value through technical means. The fact that this article has no clear answers to the low-level issues does not compromise its contribution, since in many, if not most, occasions, there is simply no clear answers. And highlighting the issues and stakes in the translation process (from high and mid-level to low-level) is itself an improvement to the design process.

The three major values served by free speech are knowledge, democracy, and autonomy: these are values that are widely endorsed by both scholars[4] and courts[5] as normative bases for the constitutional protection of this liberty. Speech or expression is not fundamentally important for its own sake – don't forget that speech can be good or bad. The reason why it receives special protection is that it is closely linked to the three values. Such intellectual consensus is a common denominator upon which the design of the networks can be based. The consensus upon the constitutional values of free speech cannot be easily transplanted to the legal rules of free speech, since it is highly debatable what these rules should be, and whether they are applicable to non-state entities (like platforms or servers in the social networks). In any case, we cannot require the private service providers or individual developers to abide with the same constitutional rules of free speech as those applied to the state. However, we can require those parties to adhere to the shared values of free speech in their design process.

This article argues that in designing the decentralized social networks, the values of knowledge, democracy, and autonomy must be considered, balanced, and implemented in an open, inclusive, and discursive manner. Each of these values entails some mid-level norms, which are also widely endorsed. For example, knowledge, as an epistemic value, requires that the knowledge production and dissemination to be smooth and efficient, and misinformation/fake news to be effectively detected, verified, and managed; democracy requires vivid, inclusive, and rational public debate which is vital for democratic governance; autonomy means that the individual users' choice and preference to be respected and realized, and that their personal faculties to be developed in a free manner. The next Part discusses whether and to what extent the decentralization design could facilitate the realization of those values.

## 4. Promises and pitfalls of decentralization

### 4.1. Knowledge

On centralized platforms, misinformation and fake news can easily go viral ([38], 85). Such velocity of spread is possible because all users are interconnected (mediated by the central). In decentralized networks, by contrast, misinformation would be much harder to go viral due to structural limitation: audience is fewer in each decentralized server than in the giant platform [39]. Without amplification by the platforms, it is nearly impossible for fake news to reach millions of users in days or even hours. In other words, decentralization reduces the harms of harmful speech by narrowing the scope of its reach. Epidemiology teaches us that one effective way of curbing the contagion of virus is quarantine. Here, similarly, by quarantining misinformation within the scope of decentralized servers, the structure has prevented them from spreading the whole network. Here, the decentralized social network's mitigating effect on misinformation comes not only from its reduced scale, as its number of users is not necessarily smaller than that of centralized platforms. Rather, the more important factor is the structural limit: the topology of decentralized networks means that each server or community is only connected with a very limited number of neighbours. This feature makes misinformation extremely hard to spread to the whole network, unlike the centralized ones.

Another way the centralized platforms facilitate the spread of misinformation is through personalized recommendations. Research found that the centralized platforms intentionally recommend misinformation to cater to the behavioral addictive tendencies and emotional drive of humans ([40], 55–8). The chief criterion of recommendation is the potential to increase user engagement, not the accuracy of information: thus, users would be hard to distinguish whether the content they see is accurate or false ([41], 79). Eradicating the engagement-based recommendation model, decentralized networks can avoid one driving force of the spread of misinformation. On decentralized networks, content is offered and sorted according users' own choices, rather than the algorithmic determination that serves the business model of platforms. Corporate powers would then have no comparable capacity to distort the marketplace of ideas through manipulative tactics. The structural feature of no centralized intermediary avoids the single point of control by the informational capitalism.

Even though the structural feature of decentralized networks significantly limits the reach of fake news and the potential of manipulation, we cannot say that the new technology only affects epistemic values in a positive way. By eradicating intermediaries and entrusting users, the decentralized networks rely on a democratic (or popular) approach of governance. This is problematic for knowledge production and dissemination for two reasons: (1) division of labor requires that "we must rely on a complex and mediating division of epistemic labor in public deliberation" ([42], 73); that means, we often do not evaluate the truth of a claim by ourselves, but rather assess the credibility and trustworthiness upon the source and form of the claim – who made the claim, on what occasion, and through what means ([42], 73). (2) knowledge, as justified true beliefs, depends on a social process of justification ([43], 444–5); and the justification needs far more than democratic assessment and oversight. Actually, modern production of knowledge cannot proceed without a series of conditions including education, press, universities, libraries, social norms, and social trust.

We can see why decentralized structure poses threat to epistemic values through the lens of the intermediaries. From mass media to (centralized) social media, the traditional intermediaries such as newspapers and televisions largely lost their gatekeeping role for knowledge production. People receive and digest knowledge mainly from the commercial platforms, and those platforms became the new intermediaries. Even though the new intermediaries lack the professional and ethical norms that bound the traditional institutions, they still exert some filtering for the content. They have established a set of moderation standards to ensure the quality of information as well as mechanisms of managing fake news. On the decentralized networks, however, even the platform intermediaries are gone. It is now the users themselves who conduct the task of identifying, evaluating, and disseminating information that they believe to be justified and true. Such reliance on users ignores the division of labor and the institutional basis for knowledge production: briefly speaking, ordinary users have neither time or institutional expertise to conduct the epistemic undertaking. Besides, as anonymous speakers, they do not have to care about long-term reputation as intermediaries or gatekeepers do ([44], 1838). It is true that Facebook did terribly bad in managing misinformation and curating its platform, but at least it has to appear in congressional hearings and face public pressure. How can we ensure that the decentralized small servers are held accountable?

[4] *See generally* [125,126].

[5] *See* Janowski v Poland, App. No. 25716/94, ECHR 1999-I, para. 30 ("Freedom of expression constitutes one of the essential foundations of a democratic society and one of the basic conditions for its progress and for each individual's self-fulfilment.")

Realizing the epistemic value of free speech is not easy. The discovery and spread of truth are only possible “under favorable demand and supply conditions” ([45], 725). Compared to the centralized platforms, the decentralized networks further diminish the role of intermediaries, putting speakers (knowledge producers) directly in front of listeners (knowledge receivers). In this circumstance, the epistemic quality of the community becomes identical to the popular preference of the members, without any filtering, editing, evaluating, and checking from external authoritative sources.

Enabling users to freely and equally determine epistemic issues for themselves simply ignores the fact that while speakers are equal, different claims of knowledge (beliefs) are not ([46], 474). Knowledge production is disciplined because knowledge claims are hierarchical – even true claims do not contain the same values in them ([43], 479). Institutions are vital to epistemic values of free speech because knowledge production requires norms, traditions, disciplines, and ethics by which knowledge are generated, scrutinized, and disseminated ([46], 481–3). They conduct the filtering task to ensure that public debate is rational and informed, and the topics for debate are reliable, relevant, and commonly referenced ([47], 151, 163, 165). We cannot expect a “naked” market without institutional clothing to perform epistemic functions well. It’s lucky that users have been liberated from the editorial tutelage of the intermediaries; it is also unlucky that they have not learnt how to make informed choices by themselves ([47], 160). “How long will it take until everyone is able to be an informed, thoughtful, and responsible author?” ([48], 848).

If a decentralized social network is based on Blockchain to store social data, an additional benefit can be gained for achieving the epistemic value. Using Blockchain can help manage misinformation or fake news, since this technology can easily identify, trace, and permanently store the source of a piece of information, facilitating efforts to evaluate the credibility of news ([49], 109; [50], 202). For example, a transparent, immutable, and certifiable registry can be generated and stored on the Blockchain, containing information about the post’s origin and credibility ([30], 2). Such enhanced transparency of information flow enabled by Blockchain can help users distinguish true and justified information from the false and unjustified, thus facilitating knowledge dissemination.

### 4.2. Democracy

Democracy is another free speech value that has been threatened by the centralized commercial platforms. Diagnosing the ills caused by the giant platforms, Fukuyama reported that “[t]heir real danger is not that they distort markets; it is that they threaten democracy.” ([51], 102)

As an alternative to the centralized platforms, decentralized networks can bring several benefits to democratic values. First, the decentralized structure is censorship-resistant ([1], 19). Even though users still have to live under the auspices of the servers’ admins or moderators (unless users host a server and moderate by themselves), the proliferation of communities makes the government or corporate powers far more difficult to exert influence on the communications ([52], 144). To censor speech on centralized social media, co-opting or collaborating with the central platform would be enough. Under the decentralized structure, such “new-school” speech regulation [53] is harder, because there is no single point of control here. In a world in which numerous communities are dispersed and loosely connected, it would be challenging to identify certain speakers and speech.

This structural feature is conducive to an inclusive speech environment, especially for social activists and marginalized groups who need a safe space for communication and community-building. But this feature also brings challenges for content moderation, as the decentralization structure makes identification of bad content harder. And effective moderation entails effective coordination across servers, thus generating risks of recentralization. Next Part of this article will elaborate this point in detail.

The second benefit from decentralization is its escape from the engagement-focus business model. Centralized platforms derive profits from user engagement, thus prefer to recommend sensational, emotional, and psychologically additive content to users; what has been sacrificed is quality, diversity and rationality of the users’ exposure to content ([54], 8–13). Such business model caused the misalignment of incentives between the platform’s goal and the public interest ([55], 88–9). Decentralized networks simply get rid of this model. In Mastodon, for instance, feeds are chronological, non-algorithmic, and adds-free ([56], 39). There is no invisible hand that exert tremendous influence upon what has been seen and discussed in the public discourse.

Third, decentralization enables more diversity and resilience to the public discourse. Centralized platforms have applied one uniform set of rules across the whole platform. This uniformity would surely be either under-inclusive or over-inclusive for different communities in the global and culturally heterogeneous platform ([13], 228). Without uniform rule enforcement by the central, decentralized networks would lower the stakes of moderation and offer more respect to cultural diversity ([52], 121–2, 157). Conflicts among divergent speech norms could be avoided by granting autonomy to each community.

Decentralized structure also avoids single point of failure (just as it avoids single point of control in resisting censorship), making the communications more resilient ([52], 122). In decentralized networks, discussion will focus on topics that are more relevant to local communities and cultures ([41], 81) – discussion in this context would be more intimate, and vivid. In the small communities, members could have stronger sense of belonging. And moderators can be more efficient in maintaining the health of the space because of more local knowledge and higher moderator-user ratio. Besides, giving individual users more chances to participate in community governance can help them develop skills and faculties that are essential for citizens to engage in democratic self-governance ([57], 7). Simply put, the online communities are democracy “schools” for citizens to practice self-governance. This stands in contrast with users in centralized platforms, who are mere passive consumers, rather than active citizen participants.

Again, decentralization’s prospect for democracy is mixed: the new project also brings threats to democracy. One of the most pressing concerns is the filter bubble effect it may exacerbate. In the age of mass media, what people can see is not determined by themselves, but by newspaper editors or TV producers. When evaluating the transformation from mass media to social media, Professor Volokh bemoaned that the common culture based on the shared information is crucial for public debate, and personalized recommendations used by social media shattered such basis so that people will find themselves having fewer shared cultural referents and common knowledge ([44], 1835). Professor Sunstein’s work proved Volokh’s prediction, that social media has facilitated the filter bubble or echo chamber effect, which harms both the common informational base for discussion and the virtue of tolerance that is vital for democracy.

Unfortunately, the centralized social media may not be the final facilitator to parochial visions and self-entrenched biases. It is true that centralized platforms use personalized recommendations to cater to users’ preferences. But they also expose to users some content that might not fit with the users’ prior preferences but the algorithms “think” may interest the users. In other words, users would still be constantly exposed to topics and viewpoints they didn’t select or even didn’t expect. Decentralization project, however, will largely terminate the appearance of such unexpected encounter. Users in decentralized networks can choose servers and moderation services at their will. More user choices guaranteed by decentralization may mean more entrenched beliefs and less exposure to challenges. Filter bubbles become more isolated here.

Writing in 1995, when the Internet was just beginning to rise and the age of social media was yet to come, Cass Sunstein imagined a “future” world that basically fits with today's reality: a world in which "each person could design his own communications universe. Each person could see those things that he wanted to see, and only these things." ([58], 1786) To Sunstein, public discourse will be democratized in such "future", but informed citizenry would be harmed, because confronting with content that people dislike can teach them tolerance and empathy, which are crucial for a functioning democracy. Democracy needs citizens to make informed choices, and for choices to be informed, citizens must be exposed to and exchange with different views and opinions, rather than rely upon information that only conforms to their preexisting preferences. Preferences without critical challenge are likely to be biased, parochial, or adaptive.

The replacement of one single platform with numerous communities not only risks entrenching the users' biases and hampers empathetic understanding. It also balkanizes the public sphere, making it less likely to host discussions nationwide or worldwide. Due to structural limitation (or, strength, depending on your position), there is no truly global view on the decentralized networks; rather, all users can only see what their servers allow them to see [4]. For free speech to facilitate democratic governance, dialogic interactions in small communities are not enough; rather, the discussion must form public opinions that influence the governance at various levels – local, national, and global. For democratic deliberation to be functioning, the numerous deliberative spaces must be able to be connected to generate meaningful dialogues that shape the national governance agenda. In the decentralized world, the public sphere will be divided into "public sphericules" ([59], 10), and voices from the dispersed servers are extremely hard to reach national audience. Of course, members of each community may govern their own communities well; but governance of a country requires opinions from different communities to engage, compete, and deliberate in an open manner. This is not what the decentralized network is good at.

Social media is not only crucial for national democratic deliberation and consensus building, but also crucial for global dialogues. Many pressing issues, such as global warming and poverty reduction, require efforts across national borders. Social media, as global platforms that do not limit within national boundaries, constitute one of the few deliberative spaces for individuals from different states and cultures to share their views on those global issues. Ensuring this space to be open, inclusive, and interconnected is thus important for the value of democracy in both national and global level. If a global public sphere is an ideal, then the giant commercial platforms are more akin to this ideal, since the decentralized networks facilitate more fragmented small spheres rather than one global big sphere.

### 4.3. Autonomy

Autonomy may be the most obvious case for decentralization to facilitate free speech values. By creating a "marketplace of filters", decentralization enables more competition and granularity ([60], 17–19), providing more choices to users. Centralized moderation has no other choice but to adopt the "one-size-fits-all" approach; however, due to cultural and personal heterogeneity, "any moderation decision is going to upset someone" ([39]b, 12). This approach then sacrifices diversity and user agency for consistency. Decentralized networks, by contrast, place the choice of who hosts users' data, what rules govern the user-generated content, and which community will users live in, directly into the hands of the users themselves. Someone described the decentralization project as a "pro-choice" movement [61]. User agency is respected to the utmost degree than any predecessors in the history of social media.

This will be the end of the story if autonomy means just more choices. But it is not, for three reasons: user autonomy comes at a price, market can also manipulate user choices, and formal choice does not necessarily guarantee substantive capabilities.

Offering abundant choices to users brings friction to the use of the social media. When you sign up on a decentralized network service, you have to choose a server to host your data; and you also need to select moderation and curation services (filters) if they are detached from servers. You will need to read the data policies and moderation policies. Apart from those, you may be forced to make choices regarding to the requests of followers or following others, whether to be included in the search function, and what is your preference of federating with others in the network. Many users may find these complex or even annoying. And the barrier created by decentralization may reflect a kind of techno-elitism ([62], 3287) that excludes ordinary users from this new project.

To some extent, the autonomy facilitated by decentralization is not given to the individual users, but the market. The market of servers, filters, and curation services constitutes the basis for users to choose. Like the economic market, such market can also fail. For instance, services and resources may concentrate to a few providers, making them monopolies that wield the power to manipulate user choices; the market may also undersupply services for users, crippling the effect of user choices. In other words, the well-functioning of the market is the precondition for user autonomy, and such precondition cannot be taken as a given.

Both the techno-elitist feature of the decentralized design and the vulnerability of the market (of servers and filters) demonstrate risks of exclusion. Even though the decentralization project aims to share power with users and include each user into the governance, the formal autonomy itself cannot ensure that the exclusionary effect of techno-elitism and market failure would not become true. Indeed, there exists a tremendous gap between formal autonomy and actual capability or power for users. More in-depth analysis on this aspect will be provided in Part 6 of this article.

Another factor that affects user autonomy is the filter bubble effect. Autonomy not only means free choice, but also entails free development of one's personalities and faculties ([63], 459). The key is, this is a dialogic process. Without insights and feedback from others, the development of our faculties and the formation of our beliefs would easily tend to be parochial and restricted ([64], 293). In small and closely related communities, people may be more willing to share their thoughts and opinions, facilitating the formation of autonomous personality. But the downside is also obvious: when "audience…are able to avoid speech they disagree with entirely, thus limiting their engagement with ideas in the market" ([65], 978), they will lose the chance to be widely exposed and critically evaluated. Exposure to diverse information, especially to those that we dislike, is critical to both collective discussion in a democratic society and development of individual faculties.

The above examination reveals that all the three values of free speech can be simultaneously promoted and compromised by the decentralized networks. On the one hand, the values themselves are abstract, complex, and contains abundant and sometimes conflicting norms. On the other hand, more importantly, the decentralization project is not a simple bricolage, but a systematic reconstruction that includes many design choices. Each design choice is a response to the value tensions that elaborated in this Part. Most value tensions can be summarized into two dilemmas in the decentralization project: the phantom of centralization and the illusion of control. These two dilemmas divide most design choices in the project, and will be introduced in the next two Parts.

## 5. The phantom of centralization

The first dilemma of the decentralization project is that it faces constant risk or lure of (re)centralization. It is a dilemma not because the bad actors will try to regain concentrated power and capture the network, making it into walled gardens again. The dilemma derives not from the malice of some players, but from the fact that there exists continuous need for the network to centralize. The need comes from three sources: epistemic management, content moderation, and cross-community discussion. Those are needed because of the epistemic and democratic values that a communications network does aim to fulfill. That's why I use "phantom" to describe the lingering force. In other words, even though we decentralize the structure of a network, it will necessarily require some centralized design for it to function well and to serve the normative values properly. Designers must then carefully consider those needs and strike delicate balance between decentralization and centralization.

### 5.1. Epistemic management

One important mechanism for ensuring the epistemic function of free speech is the social identification of experts, or, "epistemic authorities, individuals, or groups to whom others defer as reliable sources of true belief" ([43], 484). Source identification and identity management, then, are crucial for the evaluation of the credibility and quality of information. Such evaluation is the basis for knowledge production and dissemination. In the age of mass media, people trust prestigious newspapers and reporters because of the professional ethics that bind them and the reputation they earned from their past records ([66], 706).

In the age of centralized social media, the platforms established a set of mechanisms to fight misinformation and promote quality reporting. They have replaced the traditional gatekeepers and act as gatekeepers themselves ([48], 858). Of course, they are often accomplices of the spread of fake news for more user engagement. But as big corporations, they face public pressure and they care about their brands. In addition, journalists and experts can get their accounts verified by the platforms so that they be trusted by other users as more reliable sources on their field of expertise. On decentralized networks, however, can we expect the servers, especially the small ones, to do the verification work? As identity information is not shared by the whole network, how can we ensure that the verification data be safely and smoothly accessed by other servers? Or, shall we resort to centralized mechanisms again?

One decentralization project, Articonf, uses the so-called "crowd journalism" to deal with fake news: the method is to rely on user voting and rating to evaluate the trustworthiness of news ([67], 3). This approach mistakenly equates epistemic with democratic values. Epistemic values cannot be fulfilled through popular means, as professional, institutional, and ethical guarantees are necessary to resolving the question of what is true.[6] Put it in another way, experts must play a role in the identification, verification, and dissemination of knowledge in a society. In this regard, the epistemic function of experts (in institutional and professional forms) unavoidably exerts some centralized influence over the communicative sphere on the decentralized networks. The professional factcheckers and news-producers will become new intermediaries ([50], 200). And designers of the project face a choice of either sacrificing the epistemic value for preserving decentralization ideal, or introducing centralized mechanisms to manage the epistemic information of the communications, including, but not limited to, the identify verification of expert speakers and the source tracking of information.

### 5.2. Content moderation

Content moderation is the management of user content according to rules of the community: it includes content removal, amplification or reduction of spread, recommendation, and other enforcement methods that affect the visibility and reach of content ([68], 2). Moderation need is perhaps the most powerful force that drives the structure of a network into centralized control. For example, starting several years ago, Reddit began to change its hands-off approach and strengthen centralized moderation to deal with the growing salience of trolls and harassment in the sub-Reddits [69].

Decentralized networks also face the challenge of dealing with bad content. This challenge is thornier for decentralized networks because their structural feature makes the action against malicious actors uncoordinated, inconsistent, and inefficient. When the far-right community Gab migrated to Mastodon, the incident generated huge concern about whether the decentralized web is resilient enough against these extremist users. One author documented that on Mastodon and the Fediverse, many servers have no moderation at all, and most are not able to effectively conduct inter-service moderation ([70], 5). Similarly, IFTAS's report showed that only 17 % servers offer 24-hour moderation coverage, and most moderators are unpaid ([71], 3).

Moderation requires three things: expertise, time, and money. Moderators in decentralized communities may have more local contextual knowledge than the those in centralized platforms ([72], 367), but expertise requires more than that. Effective moderation requires basic knowhows about federation and protocols ([8], 1199), as well as skills of developing or utilizing technical tools to scale the moderation efforts. Moderation is costly; unlike centralized platforms which derive the funding of moderation from surveillance capitalism ([73], 981–8; [13], 221), admins and moderators on the decentralized networks have much less resources ([5], 2) and no sustainable source of funding so far. There is currently no empirical study on the exact budgetary cost for moderating a server; one author reported that he spent about 30 U.S. dollars a month to maintain a 50-people server, and it really takes much time to do the job [74].

Mike Masnick argued that "content moderation at scale is impossible to do well" [11]. The logical step that he missed is that content moderation at scale surely leads to centralization, and centralized moderation is likely to fail because of the problems of the one-size-fits-all approach, unchecked power, and manipulation, all you can see from Facebook in recent years. Even though decentralization moves power from the platforms to the users, the resource-intensive endeavor of content moderation may finally concentrate into the holds of a few who have the resources to do the work [75].

For this reason, Rozenshtein argued that the need of moderation may generate a centripetal force that recentralize the Fediverse ([13], 230). His view was echoed by other writers ([76], 7). In some sense, recentralization may be natural resort for self-defense in front of the burgeoning of bad content in the decentralized network, especially when it begins to scale in the future [77]. In one recent survey, most interviewed developers and admins agreed that "some degree of centralization is required for effective moderation" ([78], 16).

In the decentralized, recentralization may be most likely come from the effort of coordination on moderation. To decrease the cost of moderation and make it more affordable, it is imperative to coordinate the use of moderation tools and strategies across servers. This has already been used by the sub-Reddit moderators but the lack of a central repository of tools still limits the effectiveness of the coordination ([79], 12–13).

We can, of course, argue that platform-wide moderation in the decentralized networks is not only unachievable [80], but also unnecessary. In any case, decentralization gives the power and freedom to moderate to users and the servers they choose to live in, and if some servers or users choose to tolerate hatred, harassment, or extremism, then it is their choice that should be respected. But the problem is, people expect the moderation service to be in the same standard in decentralized networks as that in the centralized platforms ([81], 4).[7] For example, many users migrate to Reddit from the giant commercial platforms because there is more user control in the sub-Reddits. But the lack of moderation support from the central in the sub-Reddits has made many users leave the space ([10], 11).

[6] *See* [113], 1272 (arguing that free speech values rely upon institutions to produce, curate, and disseminate content according to professional and public-regarding norms).

[7] *See also* [71] (arguing that users may want a better moderated Twitter, rather than a self-governed community).

After Fukuyama and his collaborators proposed the Middleware plan, the Journal of Democracy organized a symposium to comment on this proposal. One of the chief concerns raised by commentators was the cost of moderation and the corresponding risk of re-centralization. For example, one author pointed out that "[m]oderating and curating content in the public interest is difficult, contentious, and expensive" ([82], 161). Another argued that because small organizations and individuals have insufficient resources and capacities to moderate, it is likely that the services will, again, be recentralized and concentrated into a few big players ([83], 164).

In designing potential coordination mechanisms of moderation, the decentralized networks must balance the need of effective moderation with the necessity of avoiding too much concentration. Questions abound: whether and how the filter or moderation tools should be shared across servers? How should they be displayed? Should there be "officially recommended" filters? Should all filters be equally listed, or should the extreme filters be excluded? ([84], 4) Who determines (in other words, who is the filter of filters)?

Another way of recentralization out of moderation need is commercial capture: because the commercial giants have more resources and experiences of dealing with bad content, their participation in the decentralized web may attract more users than the small servers, resulting in business capture of the space as well as new walled gardens ([10], 15). One illustrative example is the commercial capture and recentralization of email ([10], 7). Because small providers cannot afford the costs of quality filtering service, especially dealing with spam, most email accounts today have been hosted by a few giant service providers (such as Gmail, iCloud, and Outlook), even though the structure of email still remains protocol-based. In the decentralized networks, the fear of commercial capture is one major concern behind the debate over whether to federate with Threads (see Part 7 of this article).

Recentralization emerged out of moderation needs may compromise free speech values. That's because free speech has an anti-majoritarian spirit that aims to protect dissident and marginal speakers. If coordination among moderators of different servers is comprehensive enough, then it is highly likely that a super moderator (or moderation rule) would emerge out of market/popular demands. This is worrisome, since "[t]he whole point of placing the freedom of speech beyond the reach of democratic politics is, after all, to prevent censorship by popular demand" ([85], 1385).

### 5.3. Cross-community discussion

The previous Part of this article argued that decentralized networks may make the public sphere more fragmented than the centralized counterpart. To counter this trend, cross-community discussion is necessary. However, federation across servers would compromise the independence of each server, especially the small and marginal ones. And the higher the degree of cross-server federation is, the more likely the network would recentralize again like a monolithic platform.

Decentralized networks, by facilitating the prosperity of small and closely-knitted communities, provide havens to marginalized groups for them to associate. Actually, the Fediverse has become homes for many LGBTQ communities: the queers left mainstream social media to find a safer place to communicate [86]. These counter publics enabled marginalized voices to "explore and nurture, before explicit engagement with and challenge to dominant discourse in 'mainstream' arenas" ([87], 837). However, it is also regrettable that marginalized voices can only be heard in small and closed communities. For them to influence the public opinion and the national agenda, they must have the chance to be found and heard by wider reach of audiences. Otherwise, the marginal will always be marginal, losing the odds of challenging and negotiating with the mainstream. One author complained that one "crucial mistake of social media was trying to force people with wildly incompatible views to co-exist in the same place" [88]. But coexisting with incompatible views and people is what exactly democracy means.

The need of federation, just like the need of moderation, is another force that may recentralize the network. When more servers or nodes are connected with each other, powerful servers could exert more influence toward other servers, and the distinction of resources owned by different servers would be more explicitly revealed.

Designers of the decentralized project must make tradeoffs between two scenarios here: each is backed up by different value considerations. On one side, it is the independence and autonomy of the communities, meaning less vulnerability of marginalized groups and less risks of recentralization and capture; on the other side, it is the connected and open discussion across communications, meaning more interactions among different groups, but at the same time also bringing more risks of recentralization.

This Part introduces three cases for the seemingly inevitable recentralization of the decentralized social networks. Even though more empirical research is needed, some preliminary findings have revealed the tendency of recentralization on Mastodon: for example, researchers found that 5 % of instances (servers) host 90.6 % users and 94.8 % content ([25], 4). This is also the case for the infrastructural level: most instances are built upon the services provided by only a few hosting providers ([25], 3). This is understandable because when decentralized networks began to scale, the pressure of epistemic management, quality moderation, and cross-community discourse will generate strong force of recentralization. Of course, we do not need to be a pessimist, holding that decentralization is only workable when the network is small and it will doom when it scales [24]. Rather, the issue is a matter of degree: if recentralization is inevitable, what degree and form of recentralization should be accepted?

## 6. The illusion of autonomy/control

User autonomy and choice constitute the prime rationale for the decentralization project. By flattening the structure and getting rid of the central node, the decentralized networks aim to empower the users and make them the governors of the space. However, the proliferation of choices and the grant of formal autonomy do not necessarily result in real power of governance. And the autonomy or control might be illusory, for the following reasons.

First, freedom of choice can sometimes lead to exclusion and lack of diversity ([9], 3), especially when such freedom is built upon the possession of technical expertise and economic resources. Just as the exclusion of marginalized and powerless groups can be reproduced from offline to the online world; the same process can also be copied from the centralized platforms to the decentralized networks. The previous Part has discussed the techno-elitism inherent in the user-choice paradigm. One of the primary tasks for the designers of the decentralized networks is to figure out more user-friendly interfaces, steps, and notices. Formal choices must be accompanied by substantive capacities to choose, so that the "pro-choice" movement would be truly empowering, rather than exclusionary.

Second, the control/autonomy paradigm puts unquestionable trust upon the value of individual choice and the self-regulating potential of market, ignoring – and even entrenching – the way power shapes choice and market ([89], 38). Market, however, is not immune from inequalities, capture, and manipulation ([90], 69). The decentralization project still falls within the neoliberal thinking, for it uses formal equality of choice to legitimize the substantive inequality of power ([89], 65). Without redesigning the power structure and empowering the individual users, formal freedom and choice cannot really change the scenario under the centralized platforms, but only legitimizes the status quo ([91], 66). One reason why there is always recentralization force in the decentralized networks is that formal structural change is not enough to change the power dynamics as well as challenge the hegemonial control exerted by the central powers.

The illusion of control brought by decentralization illustrates a type of tension. The project reflects a libertarian ideal [92] that relies on the spontaneous collective action by users themselves – through technical coordination, market mechanisms, and social collaboration. It also brings tremendous burdens, frictions, responsibilities, and non-minimal vulnerabilities to malicious actors and content. In the end, it seems that either the decentralized web will scale to an extent that invite a high degree of recentralization, or it will keep its decentralized ethos but limited to a circle pf very few people that are truly capable of exercising the autonomous choices [92].

Such illusion echoes the consent dilemma in the privacy law literature. Just like the consent mechanism in privacy law, decentralization can also be utilized as a legitimation strategy. Consent has become the cornerstone of data protection laws in the world. Take EU's GDPR as an example, consent from data subjects is one of the major legitimate bases for data processing. However, because users do not have the time, expertise, and resource,[8] the consent mechanism has become mere practice of box-ticking ([93], 189). Data subjects rarely read the long and complex privacy policies. This reality has made privacy scholars doubtful about the practical utility of consent.

Likewise, most users do not read the information screen of Mastodon, and many of them find the step of choosing servers too complicated ([94], 16). The effective exercise of user agency depends on users being capable of doing so: this is not the case for the current cyberspace [95]. Indeed, more choices for users may mean less frictionless experience ([23], 159): it is debatable which stands closer to autonomy. Actually, users have complained that Mastodon is confusing and hard to see, and some have left the network for exactly this reason [96].

When the former CEO of Twitter, Jack Dorsey, initiated BlueSky, someone doubt that his motive was to foist the responsibility of moderating bad content onto the users/communities [97]. When users are unprepared and incapacitated to fulfill the task, the decentralization project would generate more burdens than choices for them, creating an illusion of control or a theatre of autonomy.

The lesson we could learn from the illusion of control is that merely proffering users more choices is not sufficient for improving user autonomy. True autonomy entails not only the capacity of having formal choices; we should also "ask which options/choices are really open to people, and how easy it is to see and understand those other options/choices" ([32], 194). To achieve true autonomy for users, both substantive and procedural empowerment is needed. For example, digital literacy projects that teach the technical knowhow to users and more friendly user interface design that smooth the use of the networks are useful methods ([23], 159). It is true that the existence of choice is itself meaningful even though users do not exercise that choice [61]. But it is also true that the mere existence of choice is not the decentralization project promises, nor what the developers, entrepreneurs, and users expect. This March, a seminar in Princeton has touched upon the topic of how to design the consent mechanism, in order to ensure that user preferences are truly respected [98]. BlueSky tried to facilitate user control in moderation by providing an easy-to-use moderating tool (Ozone) to users [99]. These are valuable endeavors that try to explore ways of turning the illusion of autonomy into realities.

One shortcoming of those current endeavors is that they are mainly discussed and decided within the circle of the technology people, lacking systematic inputs from policymakers, social scientists, and most importantly, ordinary users. One of the lessons we've learnt from centralized social media is that the basic design choices should not be made by the few powerful, whether they are technical elites, corporate tycoons, or political figures. If the key design is still discussed and made within a small circle, then what is the real difference between the decentralized web and the proprietary one anyway? An open and democratic web means not only open and democratic usage, but more importantly, open and wide participation in its governance.

Thus, to truly empower the users, we should supplement the substantive value-based design with the procedural elements. In other words, user inputs should be solicited and respected during, not after, the design process. The biggest dilemma for the consent mechanism in privacy law is that users cannot negotiate the terms due to lack of bargaining power ([91], 66). Formally, they are free to choose. But in reality, they have no alternatives but to click "agree". This should remind the designers of the decentralized networks that user choice should be provided during the design process, rather than after the design choices have been made by the few. When the design process is finished, many choices offered to the users are merely formal, because many options were excluded in the design process. Choice-after-design would make users stuck in the same dilemma as in the privacy law. The key issue behind the design choice is legitimacy, and legitimate requires that the process must open, inclusive, and discursive.

## 7. Design examples

The phantom of centralization and the illusion of autonomy reveal that there exist conflicting views about how to best achieve the design values as well as how to resolve tensions between different values. It is understandable, just as we can universally agree upon the normative values of free speech but are still highly divided on the issues in specific cases. When we move from abstract design values to specific design requirements, choices, balances, and tradeoffs must be made. This Part offers several typical design examples that have been recently discussed in the decentralized networks. During the discussions, some advocates are more prone to the decentralization ideal, preferring small, safe and intimate communities rather than coordinated control and big speech forums. Others are more tolerant to the risk of recentralization, providing more frictionless user experience and more coordination for moderation and cross-community communications.

### 7.1. Bundling or detaching

Bundling or detaching is the design choice that divides the two most popular decentralized networks to date: Mastodon and BlueSky. Functions of data hosting and content moderation is bundled in the former while separated in the latter.[9] In other words, Mastodon servers not only host the data of users, but also moderate their content; in this case, the selection of servers is the single most important thing for the users. By contrast, through decoupling of hosting and moderation, BlueSky makes the switch of moderation services easier and the selection of services less consequential ([100], 1).

Theoretically, lower switching threshold makes the network more resistant to recentralization. First, there may exist a lock-in effect on the servers of Mastodon because users have to find a new home for their data if they are only dissatisfied with the moderation, not the data hosting. And subscribing to a new filter is much easier than migrating all one's data to a new server or hosting a server by one's own ([100], 4–5, 8). Second, bundling of hosting and moderation into one layer (the server) may make corporate capture easier since the points of control are fewer than the scenario in which the two services are separated [101]. For these concerns, Masnick has commented that the over-reliance on server admins means that Mastodon users are just moving from a big centralized network (such as Facebook) to a bunch of small centralized networks (controlled by those admins) [102]. BlueSky, by detaching more layers, achieves higher degree of decentralization and avoids single point of failure.

However, the Mastodon approach is not without merit. Requiring

[8] [91], 67 (arguing that "Individuals are mostly too time-poor, resource-poor, and lacking in the necessary expertise to meaningfully make use of these individual rights.")

[9] [127] (noting that in the AP protocol, PDS stores user data while does not engage in content moderation).

data servers to conduct content moderation makes it easier for moderation to be coordinated. The servers host user data, so they have access to those data. Such access facilitates the accuracy and efficiency of moderation. The BlueSky approach, by outsourcing more points of control, also creates more points of risks. The third-party moderation services (filters) have to figure out ways to coordinate with the data servers on both content moderation and data security.

### *7.2. Default server or not*

Previously, when users sign up on the Mastodon network, they have to select a server from a page of server names. There was no recommended server to the user. This has caused some inconveniences for some users. Recently, Mastodon has provided a default server for users in their sign-up process: *mastodon.social*, the server hosted by the founder. Similarly, BlueSky users will first sign up to the main server if they do not change default; then they can choose to switch to other data stores and subscribe to other moderation services if they want. This sign-up process is as easy as that of Twitter [103].

The benefit of this design lie is twofold. First, it will surely ease the sign-up process, especially for ordinary user who find the server selection procedure complicated. This will make the decentralized networks more accessible and less techno-elitist. Second, smooth sign-up process will help scale the network, inviting more users to join. This will make the discursive sphere more open and inclusive, rather than limited to small circles.

However, the dark sides of the default server design are also obvious. First, it will bring recentralization as many users tend to choose the default server [104]. There will be real risk that the default server would become the new central node, exerting excessive influence over the whole network. Second, user autonomy may be compromised since the default server may be misleading to some users, signaling them that the default is the best, or most suitable, or even the only server for them. Such design also grants the default server undue advantage over other servers in the market, potentially creating a monopoly that distort the market.

### *7.3. Shared blocklist*

As previously argued, coordinated moderation is one of the biggest challenges for the decentralized networks. Currently, one daunting obstacle for coordinated moderation is the lack of uniform information base. As “no one has access to platform-wide data”, “analysis of coordinated threats across servers” is impossible; as a result, there lacks “a unified evidence base for threat hunting” ([78], 16–7).

One effective way of dealing with this issue is to create a shared blocklist: this enables every instance admin to know what has been blocked by other instances, and they can easily get a sense of who are the bad actors [7]. However, as argued by Part 5, coordinated moderation efforts unavoidably bring out recentralization risks. Here, the recentralization risk lies in the management of the blocklist. First, there might emerge a few super lists that are widely adopted by the servers. This would produce a de facto uniform moderation that significantly resembles the “de jure” uniform moderation by the centralized platforms. Under such circumstance, marginalized views would find few places to hide, as most servers are following the majoritarian views expressed by those super blocklists. The communication space, unfortunately, would become more homogeneous and less diverse.

Second, how to manage the list, and how to prevent it from being abused? As a site of centralized power, the blocklist faces multiple sources of threat, such as the capture from corporate and government, as well as the abuse from malicious actors in the community. Third, how to make sure that the blocklist remains to be fair and transparent? For example, people complained that The Bad Space (one popular shared blocklist) does not allow servers the chance to appeal, and some even accused that the list has been abused to silence the voice of the LGBTQ groups [105]. Do users and moderators have due process rights (such as the right to appeal) regarding the list?

### *7.4. Opt-in or opt-out federation*

Federation means to build connections among different servers/communities in the decentralized network. Mastodon has adopted the approach of opt-out federation among its instances. That means, all federation requests from other instances will be automatically approved, unless the admins choose to opt out of this default setting. This raises concern among admins and moderators about the risk of unexpected bad content. Some have advocated for changing the federation to consent-based (or opt-in) ([12], 161). The rationale behind is that easier federation will increase the difficulty of moderation, since malicious servers will encounter little friction in connecting with others. But Mastodon insisted that the default federation conforms to its goal of creating an open web [106].

Apart from the default setting of federation, there are two related design choices: whether the Fediverse servers (using ActivityPub protocol) should federate with BlueSky (using AT protocol), and whether the decentralized networks should federate with centralized commercial ones like Threads. When being asked about whether BlueSky plans to interoperate with the Fediverse, BlueSky chief Graber said its currently not on schedule but the team does not oppose this idea [107]. Someone has already started working on the bridging tools between the two networks [108] but the project’s core issue is, again, whether the bridge should be opt-in or opt-out [86]. Opt-in enables more user choice while opt-out encourages easier and quicker federation across protocols [109]. Recently, the developer of BridgyFed, Barrett, changed the federation from opt-out to opt-in for the respect of user autonomy [110]. However, more autonomy in this case also means less federation and lower rate of cross-community communication. It is true that users should have the say on the scope their audience. But is also true that we constantly encounter unexpected messages and unexpected audiences in our daily lives (both offline and online) and such experience is often enriching for both personal development and public discussion.

With regard to Threads, it seems that the Fediverse community is highly divided on whether to federate with that giant behemoth, reflecting different visions of the future of the network [111]. Those who oppose federation fear the scale of users would defy effective moderation, and Threads will bring centralization and surveillance capitalism to the network again. Those who support federation, by contrast, prefer bigger and more open communities rather than small and isolated ones [70,111].

How to design the federation mechanism of the communities on the network? Opt-in or opt-out? This is a choice that reflects different normative visions regarding the small versus the big community [112], or, different value commitments. Lying on the small community side are values of personal autonomy and incubating spaces for both knowledge production and democratic discussion. On the one hand, consent-based federation respects user choice. On the other hand, “[c]irculation of informed opinions often requires breathing space for reflection, consideration, and opinion formation. Knowledge takes time to absorb and assimilate” ([113], 1264). Safe and intimate discussions in a sheltered space are often crucial for the development of ideas before they are ready to be shared in the public sphere ([114], 476). Lying on the big community side are values of openness, cross-community exchange, and consensus building on wider bases. Knowledge needs critical assessment for it to be justified, and democratic discourse also has the goal of connecting widespread political communities and shaping national consensus. Designing federation schemes should carefully consider those concerns on the two sides.

### *7.5. Opt-in or opt-out search*

The search function has been another fiercely debated issue in the decentralized circle. For several years after its launch, Mastodon did not develop the function of global search mainly out the insistence on the decentralization ideal [115]. This design choice has costs: poor discoverability constitutes one of the major reasons why people left Mastodon [96]. Recently, Mastodon has changed its mind and added search to its application. The search function incorporated by Mastodon is opt-in: that means, content can only be indexed into the search after user consent. The problem is, few users change the default setting. Therefore, the opt-in approach has resulted in too few people participated in the search – <5 % after the launch of the search function, [116]. This greatly compromises the value of search since it depends on a broad coverage of content.

The debate over search mirrors the tension between decentralization and centralization. The function of search enables content to be citable and referenceable over time: this enables knowledge production and consensus building across communities/servers [117]. However, search also increases visibility and vulnerability, especially for marginalized users whose presence may be easily identified by malicious actors; it also facilitates data surveillance and privacy encroachment by those who are not the intended audience of the original speaker [117]. On the one hand, search serves important values like collective discussion, agenda setting, and consensus building across communities. On the other hand, it may encroach upon the isolated shelters for marginalized people. In addition, the global view on the whole networks as well as the comprehensive database generated by the search function may become crucial factors that drive recentralization, facilitating capture or control by some powerful actors on the networks.

### *7.6. Federated timeline*

Federated timeline refers to the column or feeds list that shows content from other servers. It is one effective way of countering the filter bubbles; by enabling users to see outside content, the timeline offers a "worldview" that includes something unexpected ([25], 2). The controversial issue is how to display the content on the timeline. Chronological? Random? Or to recommend some content that is contrary to users' preference or ideological position?

Here, we see again the division between Mastodon and BlueSky. Mastodon only allows chronological order of posts, while BlueSky allows users to choose curation algorithms from the market so that their preferences could be fulfilled in the "worldview" ([100], 3). What's more, the global view over the network is more user-friendly in BlueSky: for instance, users could see "global" posts with all replies, regardless of whether the replies are from the same server of the user ([100], 7). Of course, algorithmic recommendation will bring the problem of centralized power. And the market of curation algorithms endorsed by BlueSky may produce concentration and manipulation. The function of "global" posts, for instance, requires an indexing structure ([100], 7), which is centralized in nature.

In contrast, Mastodon's algorithmic-free approach avoids centralization risk to the greatest, but it invites other problems. First, users would be hard to sort content from the timeline and choose what really interests them. Refusing to provide other options other than chronological sorting dampens the users' choice. Second, "chronological sorting strongly favors active posters over inactive ones, without considering the signal/noise ration of those posts" [118].

We can see from these design examples the two dominant issues of the decentralized web: how to balance the decentralization goal with the centralization needs, and how to empower users to enable them better exercising their choices in an autonomous way. The normative considerations behind can be recapitulated by the three value dimensions listed by this article: knowledge, democracy, and autonomy. For knowledge production and democratic discourse, we see the case for small shelters, community ethos, and safe discursive spaces; we also see the case for open dialogue, cross fertilization, and broad participation. The decisions made regarding these design issues will determine the future of the decentralized networks. And they cannot be made in some cursory or ignorant ways. One contribution of this article is to underscore the value tensions and considerations behind these choices, shedding new light on the debate about them.

## 8. Law and policy interventions

To make those choices of design mentioned in the previous part, cross-disciplinary discussion is surely necessary. What's also needed is proper interventions from law and policy. The interventions could serve as important incentives, the aim of which is to ensure that values are seriously considered and tradeoffs are prudently made in the design process of the decentralized networks. So far, lawmakers and regulators in the world have paid little attention yet to regulating this new technical paradigm ([76], 8). This part offers some tentative thoughts on how law and policy could facilitate the value-based design in this area. These thoughts are far from conclusive and aim to invite more proposals in the future.

### *8.1. Use liability scheme to balance decentralization and centralization*

Moderation comes with liability. Currently, the law of liability for decentralized networks is unclear. No jurisdiction has passed rules for this new type of network. And the liability rules for centralized platforms cannot be directly applied to the decentralized networks, because they have distinctive incentive structure, technical capacity, and economic resources as compared with their centralized counterparts. They lack both incentives and resources to do high-quality moderation, so the liability rule for them should be different from that for the commercial giants ([119], 24). In designing liability rules for them, the factors of incentives and resources should be considered. What's more, the liability rule can also be used as a lever to balance two competing values: decentralization and centralization.

As argued by this article, moderation is a driving force toward centralization. More moderation liabilities provide an incentive for more coordination efforts and more moderation tools, creating centripetal forces among the servers. Regulators could use conditional liability as a tool to adjust the degree of centralization in the network. For example, if malicious and unlawful content are flooding on the network, regulators could impose stricter liability rules into the server admins so that more coordinated and centralized efforts can be made to combat those content; and vice versa, if the degree of centralization is too high, looser liability rules can be devised.

### *8.2. Use interoperability rule to balance federation and defederation*

EU's Digital Market Act prescribes the interoperability requirement for messaging services [120]. Social media platforms were initially included but then deleted from the final draft of the law [121]. Interoperability rules can play important roles for the development of the decentralized networks. Legal rules could intervene by requiring interoperability under truly equitable terms: "government could mandate that large social media platforms interoperate with the Fediverse" ([13], 234), and centralized platforms could not discriminate decentralized network servers or third-party service providers ([122], 1016). The law can impose requirement upon the API of the centralized platforms ([21], 108) so that users could easily migrate to or connect with the third-party moderation service providers and decentralized servers. In foreseeable future, decentralized networks will co-exist with centralized platforms. They may have different, though sometimes overlapping, user bases. Ensuring that the two different networks can connect and federate with each other is important: this is not only the spirit of an open Internet, but also facilitates mutual learning and innovation.

Interoperability rules are not only crucial for the relationship between centralized and decentralized networks; they can also regulate the connection between servers within the decentralized networks. The former serves the purpose of reducing entry barriers, promote competition, and facilitate the development of the decentralized project. The latter can be used as a regulatory tool to balance the decentralization and federation in the network: by requiring the conditions and terms of interoperability, the law regulates how the servers in the decentralized networks can federate with each other. Granting more autonomy to communities themselves would serve the values of independence and decentralization, while mandating easier interoperability (or even setting it as the default, as what Mastodon does) can facilitate more connection and openness in the network.

### *8.3. Encourage user participation in the design process*

Another way through which law and policy could intervene is to encourage participatory design of the decentralized networks. As argued earlier, user choices after the design are sometimes illusory and even symbolic. For example, opt-in search functionality surely gives users the freedom to choose whether to join the search. But the opt-in search itself is a design choice. The choice of opt-in or opt-out, and the design details of how to opt in, including the user interface of opt-in, the convenience of making the choice, as well as relevant notices and follow-up monitoring tools, matter no less than the binary choice of opt-in or opt-out. These design choices should be made with the input from users.

To enable user choices during design, procedures and mechanisms should be built. Workshops between designers and user representatives, user surveys, and user-expert codesign are some useful forms that facilitate user participation in the value-based design process. User's input should be recorded, discussed, and responded in the discussions. The process should also be transparent to invite public oversight.

## 9. Conclusion and future work

The decentralization vision represents a road untrodden, an alternative world, and another possibility for the future of free speech online. The current landscape, which is dominated by centralized, proprietary, and commercial platforms, has witnessed a series of pathologies of our online communications: unbridled power to censor, monopoly in the market, and the surveillance-based business. The decentralized social networks have been taken as a cure, representing the future of the Internet. This article makes a critical and normative evaluation of the new project. It shows that the new alternative does have potentials for a better online space, but this should not be taken as granted. For one thing, decentralization is not inherently better or freer than its centralized counterpart; it just presents new possibilities for more freedom and more agency ([123], 172). For another, decentralization cannot save every ill of the centralized platforms, and value conflicts and tradeoffs are always inevitable.

Deconstructing the central powers not only means the empowerment of the nodes, or peripheries. It also means the increased burden on the nodes (end users) to manage the sphere, to fund the sphere, and to keep the sphere safe, vivid, and inclusive. Achieving these requires far more than just devolving the power from the central to the end. Also needed is a myriad of conditions, including but not limited to a technical design that balances decentralization and centralization, a viable business model that run the network, and a suitable legal framework that manages the relevant risks and duties. Technicians, entrepreneurs, and lawyers should work together in thinking about how to design the networks to better facilitate the values that we expect decentralization to achieve.

To further explore how the decentralization social networks impact free speech values and how the technology can be designed to better serve those values, future work of research can be done in at least three aspects. First, empirical studies should be conducted to examine the effect of different design choices on free speech values. The design choices listed in part 7 represent some of the most hotly debated issues in the decentralization circle; many choices contain both promises and risks for the online discursive space. In addition, more design options will emerge in the future. Scholars can empirically test these specific design options against the free speech values: in particular, under what conditions the design would facilitate or compromise certain values. For example, different approaches of managing the shared blocklist can be studied to evaluate their impacts on deterring misinformation on the network; and the echo-chamber effect can also be quantitatively examined under different design choices.

Second, small decentralized social networks should gain more attention in future works. Currently, most academic debate, including this article, has focused on those major (most popular) decentralized social media, such as Mastodon and BlueSky. There are numerous others that are not so popular but offer important niched communities and diverse functionalities. They worth scholarly attention as they represent new possibilities and experimentations of design in this field. Third, future studies should conduct closer examination on the relationship between the decentralized social networks and their centralized counterparts. It seems that the two types of networks will co-exist for a long time – and it is good for reasons of diversity, innovation, and competition ([26], 5). But how the two will coexist, especially the way in which they connect and communicate with each other, will significantly shape the future landscape of online social interactions. Such crucial issue awaits more research endeavors and explorations.

## Declaration of competing interest

The authors declare that they have no known competing financial interests or personal relationships that could have appeared to influence the work reported in this paper.

## Data availability

No data was used for the research described in the article.

## Acknowledgement

The work described in this paper was fully supported by a grant from City University of Hong Kong (Project No. 7200786).

## References

[1] Gehl R. Alternative social media: from critique to code. The sage handbook of social media. SAGE; 2018.
[2] Ricknell E. Freedom of Expression and Alternatives for Internet Governance: Prospects and Pitfalls. Media and Communication 2020;8(4):110–20. Oct 15.
[3] Graber J. Blockchain social networks: using blockchains for monetization and data storage [Internet]. 2020. Available from: https://medium.com/decentralized-web/blockchain-social-networks-c941fb337970.
[4] Shaw C.R. Decentralized social networks: pros and cons of the mastodon platform. 2020.
[5] Zia H, Raman A, Castro I, Hassan Anaobi I, De Cristofaro E, Sastry N, et al. Toxicity in the decentralized web and the potential for model sharing. Proc ACM Meas Anal Comput Syst 2022;6(2):1–25.
[6] Marx J., Cheong M. Decentralised social media: scoping review and future research directions [Internet]. AIS electronic library (AISeL). 2023. Available from: https://aisel.aisnet.org/acis2023/58.
[7] Cohn C, Mir R. The fediverse could be awesome (If we don't screw it up). Electronic Frontier Foundation; 2022 [Internet][cited 2024 Apr 11]. Available from: https://www.eff.org/deeplinks/2022/11/fediverse-could-be-awesome-if-we-dont-screw-it.
[8] Zulli D, Liu M, Gehl R. Rethinking the "social" in "social media": insights into topology, abstraction, and scale on the Mastodon social network. New Media Soc 2020;22(7):1188–205.
[9] Mannell K, Smith ET. Alternative social media and the complexities of a more participatory culture: a view from Scuttlebutt, 8. Soc Media Soc; 2022. p. 1–11.
[10] Struett, Thomas and Sinnreich, Aram and Aufderheide, Patricia and Gehl, Robert, Can This Platform Survive? Governance Challenges for the Fediverse (August 27, 2024). pp1-21. Available at SSRN: https://ssrn.com/abstract=4598303.
[11] Masnick M. Masnick's impossibility theorem: content moderation at scale is impossible to do well. Techdirt; 2019.

[12] Caelin D. Decentralized networks vs the trolls. Fundamental challenges to global peace and security. Cham: Springer International Publishing; 2022. p. 143–68. https://doi.org/10.1007/978-3-030-79072-1_8 [Internet]Available from:.
[13] Rozenshtein AZ. Moderating the fediverse: content moderation on distributed social media. J Free Speech Law 2023;3(1):217–36.
[14] Roscam Abbing R, Diehm C, Warreth S. Decentralised social media. Internet Policy Rev 2023;12(1):1–11.
[15] Gillespie T. Custodians of the internet: platforms, content moderation, and the hidden decisions that shape social media. Yale University Press; 2018.
[16] Obar J, Wildman S. Social media definition and the governance challenge: an introduction to the special issue. Telecomm Policy 2015;39(9):745–50.
[17] Dijck J. The culture of connectivity: a critical history of social media. Oxford University Press; 2013.
[18] Packingham v. North Carolina, 137 S. Ct. 1730 (2017).
[19] Mathew AJ. The myth of the decentralised internet. Internet Policy Rev 2016;5(3):1–16. Sep 30.
[20] Fukuyama F, Richman B, Goel A, Katz RR, Melamed AD, Schaake M. Report of the working group on platform scale. Stanford Cyber Policy Center; 2020. p. 1–45. Nov.
[21] Fukuyama F, Richman B, Goel A, Katz RR, Melamed AD, Schaake M. Middleware for dominant digital platforms: a technological solution to a threat to democracy. Stanford Cyber Policy Center; 2021. p. 1–13. Jan.
[22] Wolfram S.. Testifying at the Senate about A.I.-selected content on the internet [Internet]. Stephen wolfram writings. 2019. Available from: https://writings.stephenwolfram.com/2019/06/testifying-at-the-senate-about-a-i-selected-content-on-the-internet/.
[23] Stasi ML. Unbundling hosting and content curation on social media platforms: between opportunities and challenges. J Law Technol 2023;28(2):138–74.
[24] Guadamuz A. Whatever happened to our dream of decentralization? [Internet]. TechnoLlama. 2022. Available from: https://www.technollama.co.uk/whatever-happened-to-our-dream-of-decentralization.
[25] Raman A, Joglekar S, Cristofaro ED, Sastry N, Tyson G. Challenges in the decentralised web. In: Proceedings of the internet measurement conference. ACM; 2019. p. 217–29. https://doi.org/10.1145/3355369.3355572 [Internet]Available from:.
[26] Dhawan S, Hegelich S, Sindermann C, Montag C. *Re*-start social media, but how? Telemat Inform Rep 2022;8:1–7.
[27] Shilina S. The future of social networking: Decentralization for user empowerment, privacy, and freedom from censorship [Internet]. Medium. 2023. Available from: https://medium.com/paradigm-research/the-future-of-social-networking-decentralization-for-user-empowerment-privacy-and-freedom-from-a0a8f74790cb.
[28] Graber J. Decentralized social networks, comparing federated and peer-to-peer protocols [Internet]. 2020. Available from: https://medium.com/decentralized-web/decentralized-social-networks-e5a7a2603f53.
[29] Freni P. Fixing social media with the blockchain: make the user great again. Over the Block; 2019 [Internet][cited 2024 Sep 12]. Available from: https://medium.com/overtheblock/https-medium-com-overtheblock-fixing-social-media-with-the-blockchain-b17c914bcd3d.
[30] Hisseine MA, Chen D, Yang X. The application of blockchain in social media: a systematic literature review. Appl Sci 2022;12(13). Jun 28.
[31] Raso, F. et al. Artificial intelligence & human rights: opportunities & risks. 2018 [cited 2024 Aug 13]. Available from: https://dash.harvard.edu/handle/1/38021439.
[32] Helberger N, et al. Choice architectures in the digital economy: towards a new understanding of digital vulnerability. J Consum Policy (Dordr) 2022;45:175–200.
[33] Boehm B. Value-based software engineering. ACM SIGSOFT Softw Eng Notes 2023;28(2):1–11. Mar.
[34] Borning A, Muller M. Next steps for value sensitive design. In: Proceedings of the CHI '12: SIGCHI conference on human factors in computing systems; 2012. p. 1125–34.
[35] Bernstein MS, Christin A, Hancock JT, Hashimoto T, Jia C, Lam M, et al. Embedding societal values into social media algorithms. J Online Trust Saf 2023;2(1):1–13. Sep 21.
[36] Sanchez v. France, App No. 45581/15, ECHR (2023).
[37] Poel I. Chapter 20: translating values into design requirements. In: Michelfelder DP, McCarthy N, Goldberg DE, editors. Philosophy and engineering: reflections on practice, principles and process. Dordrecht: Springer; 2013. p. 253–65.
[38] Napoli PM. What if more speech is no longer the solution? First amendment theory meets fake news and the filter bubble. Federal Commun Law J 2018;70(1):55–104.
[39] Masnick M. Protocols, not platforms: a technological approach to free speech. Knight First Amendment Institute; 2019 [Internet]Available from: https://knightcolumbia.org/content/protocols-not-platforms-a-technological-approach-to-free-speech.
[40] Ho L. Countering personalized speech. Northwestern J Technol Intel Prop 2022;20:39–86.
[41] Lafont C. A democracy, if we can keep it. Remarks on J. Habermas' a new structural transformation of the public sphere. Constellations 2023;30(1):77–83.
[42] Festenstein M. Truth and trust in democratic epistemology. In: Geenens R, Tinnevelt R, editors. Does truth matter? Democracy and public space. Dordrecht: Springer; 2009. p. 69–79.
[43] Blocher J. Free speech and justified true belief. Harv Law Rev 2019;133(2):439–96. Dec.
[44] Volokh E. Cheap speech and what it will do. Yale Law J 1995;104(7):1805. May.
[45] Wonnell C. Truth and the marketplace of ideas. UC Davis Law Rev 1986;19(3):669–728.
[46] Horwitz P. The first amendment's epistemological problem. Washington Law Rev 2012;87(2):445–94.
[47] Habermas J. Reflections and hypotheses on a further structural transformation of the political public sphere. Theory Cult Soc 2022;39(4):145–71. Jul.
[48] Susen S. A new structural transformation of the public sphere? With, against, and beyond Habermas. Society 2023;60(6):842–67. Nov 15.
[49] Al-Saqaf W, Edwardsson MP. Could blockchain save journalism? An explorative study of blockchain's potential to make journalism a more sustainable business. Blockchain and web, 30. Routledge; 2019. p. 97–113.
[50] Toumanidis L, Heartfield R, Kasnesis P, Loukas G, Patrikakis C. A prototype framework for assessing information provenance in decentralised social media: the EUNOMIA concept. In: Proceedings of the international conference on e-democracy. Cham: Springer International Publishing; 2019. p. 196–208. https://doi.org/10.1007/978-3-030-37545-4_13 [Internet]Available from:.
[51] Fukuyama F, Richman B, Goel A. How to save democracy from technology: ending big tech's information monopoly. Foreign Affairs 2021;100(1):98–110. Jan/Feb.
[52] Guggenberger N. Moderating monopolies. Berkeley Technol Law J 2023;38(1):119–71.
[53] Balkin J. Old school/new school speech regulation. Harv Law Rev 2014;127:2296.
[54] Article 19. Taming Big Tech Protecting freedom of expression through the unbundling of services, open markets, competition, and users' empowerment [Internet]. 2021. Available from: https://www.article19.org/wp-content/uploads/2021/12/Taming-big-tech_FINAL_8-Dec-1.pdf.
[55] Balkin JM. How to regulate (and not regulate) social media. J Free Speech Law 2021;1:71.
[56] Graber J. Ecosystem review. 2021 Jan; Available from: https://t.co/U5DczWX1qb.
[57] Zuckerman E, Rajendra-Nicolucci C. From community governance to customer service and back again: *re*-examining pre-web models of online governance to address platforms' crisis of legitimacy. Soc Media Soc 2023;9(3).
[58] Sunstein CR. The first amendment in cyberspace. Yale Law J 1995;104(7). May1757.
[59] Lyons BA. From code to discourse: social media and linkage mechanisms in deliberative systems. Regular Issue 2017;13(1):1–35.
[60] Allen D., Lim W., Frankel E., Simons J., Siddarth D., Weyl G. Ethics of decentralized social technologies: lessons from Web3, the fediverse, and beyond. Edmond and Lily Safra Center for Ethics. 2023 Mar.
[61] Angwin J. What if you knew what you were missing on social media? The New York Times; 2023 [Internet]Aug 17 [cited 2024 Apr 8]; Available from: https://www.nytimes.com/2023/08/17/opinion/social-media-algorithm-choice.html?unlocked_article_code=mT6ntAeiTg4MF3WsZFYZS5Gnpaxv1n.
[62] Gehl RW, Zulli D. The digital covenant: non-centralized platform governance on the mastodon social network. Inf Commun Soc 2023;26(16):3275–91.
[63] Bollinger LC. Free speech and intellectual values. Yale Law J 1983;92(3):438. Jan.
[64] Shiffrin S. A thinker-based approach to freedom of speech. Const Comment 2011;27(2):283–307.
[65] Jones R. Can you have too much of a good thing: the modern marketplace of ideas. Miss Law Rev 2018;83(4):971–88.
[66] Bimber B, Gil de Zúñiga H. The unedited public sphere. New Media Soc 2020;22(4):700–15. Apr.
[67] MOG Technologies team. Countering fake news with knowledge. Articonf; 2021 [Internet]Available from: https://articonf.eu/countering-fake-news-with-knowledge/.
[68] Gillespie T. Do not recommend? Reduction as a form of content moderation. Soc Media Soc 2022;8(3):1–13.
[69] Greenberg J. Reddit wants to exile trolls. But growing up is hard. WIRED 2015 [Internet]May 15Available from: https://www.wired.com/2015/05/reddit-wants-exile-trolls-growing-hard/.
[70] Jon. Should the fediverse welcome its new surveillance-capitalism overlords? Opinions differ! (UPDATED for 2024). The nexus of privacy [Internet]. 2023 Dec 31 [cited 2024 Apr 13]; Available from: https://privacy.thenexus.today/should-the-fediverse-welcome-surveillance-capitalism/#two-views.
[71] Hof L. Last week in fediverse – episode 39 [Internet]. The Fediverse Report. 2023 [cited 2024 Apr 9]. Available from: https://fediversereport.com/last-week-in-fediverse-episode-39/.
[72] Rosenthal H, Belmas G. Cyber-recapitulation? What online games can teach social media about content management. Jurimetrics 2021;61:331–78.
[73] Balkin JM. The first amendment in the second gilded age. Buffalo Law Rev 2018;66(5):979–1012. Dec.
[74] Kazemi D. How to run a small social network site for your friends [Internet] Run Your Own Soc 2019. Available from: https://runyourown.social/#you-get-to-makes-the-social-rules-and-policies.
[75] Pierce D. Can ActivityPub save the internet? The Verge; 2023 [Internet]Apr 20Available from: https://www.theverge.com/2023/4/20/23689570/activitypub-protocol-standard-social-network.
[76] Friedl P, Morgan J. Decentralised content moderation. Internet Policy Rev 2024;13(2):1–11. Apr 4.
[77] Lawson N. Mastodon and the challenges of abuse in a federated system [Internet] Read the Tea Leaves 2018. Available from: https://nolanlawson.com/2018/08/31/mastodon-and-the-challenges-of-abuse-in-a-federated-system/.

[78] Roth Y, Lai S. Securing federated platforms: collective risks and responses. J Online Trust Saf 2024;2(2):1–51.
[79] Jhaver S, Birman I, Gilbert E, Bruckman A. Human-machine collaboration for content regulation. ACM Trans Comput-Hum Interact 2019;26(5):1–35. Jul.
[80] Robertson A. How the biggest decentralized social network is dealing with its Nazi problem. The Verge; 2019 [Internet]Jul 13Available from: https://www.theverge.com/2019/7/12/20691957/mastodon-decentralized-social-network-gab-migration-fediverse-app-blocking.
[81] Newton C. Bluesky, threads, and the decentralization dilemma. Platformer 2023 [Internet]Jul 18Available from: https://www.platformer.news/bluesky-threads-and-the-decentralization/.
[82] Maréchal N. The future of platform power: fixing the business model. J Democr 2021;32(3):157–62. Jul.
[83] Ghosh D, Srinivasan R. The future of platform power: reining in big tech. J Democracy 2021;32(3):163–7. Jul.
[84] Esber J. et. al., A meta-proposal for Twitter's bluesky project (March 31, 2021). pp1-5. Available at SSRN: https://ssrn.com/abstract=3816729.
[85] Langvardt K. Regulating online content moderation. Georgetown Law J 2018; 106:1353–88.
[86] Barrett R. Moderate people, not code. Snarfed; 2024 [Internet][cited 2024 Apr 6] Available from: https://snarfed.org/2024-01-21_moderate-people-not-code#more-51922.
[87] Dahlberg L. Rethinking the fragmentation of the cyberpublic: from consensus to contestation. New Media Soc 2007;9(5):827–47. Oct.
[88] Hutchins M. I was wrong about Mastodon – Marcus Hutchins. 2022 Nov 30; Available from: https://marcushutchins.com/blog/tech/opinions/i-was-wrong-about-mastodon-moderation.html.
[89] Bietti E. A genealogy of digital platform regulation. Georgetown Law Technol Rev 2023;7(1):1–68.
[90] Griffin R. Public and private power in social media governance: multistakeholderism, the rule of law and democratic accountability. Transnational Legal Theory 2023;14(1):46–89.
[91] Edwards L, Veale M. Slave to the algorithm? Why a "right to an explanation" is probably not the remedy you are looking for. Duke Law Technol Rev 2017;16(1): 18–82. May 23.
[92] Heaven D. A plan to redesign the internet could make apps that no one controls. MIT Technol Rev 2020 [Internet]Jul 1Available from: https://www.technologyreview.com/2020/07/01/1004725/redesign-internet-apps-no-one-controls-data-privacy-innovation-cloud/.
[93] Alsenoy B, Kosta E, Dumortier J. Privacy notices versus informational self-determination: minding the gap. Int Rev Law Comput Technol 2014;28(2): 185–203.
[94] Laude M., Brewitz M. Centralisation on decentralised online social networks. 2023.
[95] Obar JA. Big data and the phantom public: Walter Lippmann and the fallacy of data privacy self-management. Big Data Soc 2015;2(2):1–16. Aug 20.
[96] Kissane E. Mastodon is easy and fun except when it isn't [Internet] Erin Kissane 2023 [cited 2024 Apr 13]. Available from: https://erinkissane.com/mastodon-is-easy-and-fun-except-when-it-isnt.
[97] Popper N. Twitter and Facebook want to shift power to users. or do they? The New York Times; 2019 [Internet]Dec 18; Available from: https://www.nytimes.com/2019/12/18/technology/facebook-twitter-bitcoin-blockchain.html.
[98] Monroy-Hernández A. Five themes discussed at Princeton's workshop on decentralized social media [Internet]. Freedom to Tinker. 2024 [cited 2024 Apr 13]. Available from: https://freedom-to-tinker.com/2024/03/19/five-themes-discussed-at-princetons-workshop-on-decentralized-social-media/.
[99] The Bluesky Team. Bluesky's stackable approach to moderation [Internet]. Bluesky; 2024 [citedApr 3]. Available from: https://bsky.social/about/blog/03-12-2024-stackable-moderation.
[100] Kleppmann M., Frazee P., Gold J., Graber J., Holmgren D., Ivy D., et al. Bluesky and the AT protocol: usable decentralized social media [Internet]. arXiv.org. 2024. Available from: https://arxiv.org/abs/2402.03239.
[101] MacManus R. Threads adopting activitypub makes sense, but won't be easy [Internet]. The New Stack; 2023 [cited 2024 Apr 8]. Available from: https://thenewstack.io/threads-adopting-activitypub-makes-sense-but-wont-be-easy/.
[102] Masnick M. Bluesky begins to make its decentralized vision real. Techdirt; 2024 [Internet][cited 2024 Apr 16]. Available from: https://www.techdirt.com/2024/02/26/bluesky-begins-to-make-its-decentralized-vision-real/.
[103] Masnick M. Why bluesky remains the most interesting experiment in social media, by far. Techdirt; 2024 [Internet][cited 2024 Apr 16]. Available from: https://www.techdirt.com/2024/03/27/why-bluesky-remains-the-most-interesting-experiment-in-social-media-by-far/.
[104] Hof L. The roundup – episode 17 [Internet]. The Fediverse Report. 2023 [cited 2024 Apr 8]. Available from: https://fediversereport.com/the-roundup-episode-17/.
[105] Jon. It's possible to talk about The Bad Space without being racist or anti-trans – but it's not as easy as it sounds. The nexus of privacy [Internet]. 2023 Nov 27 [cited 2024 Apr 13]; Available from: https://privacy.thenexus.today/the-bad-space/.
[106] Jon. Strategies for the free fediverses (Work in progress). The nexus of privacy [Internet]. 2024 Jan 1 [cited 2024 Apr 15]; Available from: https://privacy.thenexus.today/strategies-for-the-free-fediverses/.
[107] Heath A. Bluesky is ready to open up. The Verge; 2024 [Internet]Feb 6; Available from: https://www.theverge.com/2024/2/6/24062837/bluesky-drops-invite-system-begins-federation-at-protocol.
[108] Barrett R. *Re*-introducing bridgy fed. Snarfed; 2023 [Internet][cited 2024 Apr 2]. Available from: https://snarfed.org/2023-11-27_re-introducing-bridgy-fed.
[109] Fox K. Beyond mastodon and bluesky: toward a protocol-agnostic federation. Kye Fox; 2023 [Internet][cited 2024 Apr 11]. Available from: https://kyefox.com/2023/10/19/beyond-mastodon-and-bluesky-toward-a-protocol-agnostic-federation/.
[110] Silberling A. Bluesky and mastodon users are having a fight that could shape the next generation of social media. TechCrunch; 2024 [Internet][cited 2024 Apr 13]. Available from: https://techcrunch.com/2024/02/14/bluesky-and-mastodon-users-are-having-a-fight-that-could-shape-the-next-generation-of-social-media/.
[111] Barber G. Meta's threads could make—Or break—The fediverse. Wired 2023 [Internet]Jul 18 [cited 2024 Apr 23]; Available from: https://www.wired.com/story/metas-threads-could-make-or-break-the-fediverse/.
[112] Prodromou E. Big fedi, small fedi. Evan Prodromou's Blog; 2023 [Internet][cited 2024 Apr 4]. Available from: https://evanp.me/2023/12/26/big-fedi-small-fedi/.
[113] Balkin JM. Free speech versus the first amendment. UCLA Law Rev 2023;70: 1206–73. Apr 10.
[114] Huang T. A quadruple doctrinal framework of free speech. Columbia Hum Rights Law Rev 2022;53:467.
[115] Smith E. How mastodon search works: why mastodon search seems so unclear. Tedium; 2022 [Internet]Available from: https://midrange.tedium.co/issues/how-mastodon-search-works/.
[116] Hof L. Last week in fediverse – episode 36 [Internet]. The Fediverse Report. 2023 [cited 2024 Apr 9]. Available from: https://fediversereport.com/last-week-in-fediverse-episode-36/.
[117] Dash A. How you could build a search that the fediverse would welcome [Internet]. Anil Dash. 2023 [cited 2024 Apr 7]. Available from: https://www.anildash.com/2023/01/16/a-fediverse-search/.
[118] Unterwaditzer M. Alternative timelines in Mastodon [Internet]. Markus Unterwaditzer. 2023 [cited 2024 Apr 12]. Available from: https://unterwaditzer.net/2023/mastodon-timelines.html.
[119] Mahadeva, Nikhil, Everyone Everywhere All at Once: The Fediverse Problem (January 15, 2024). pp1-46. Available at SSRN: https://ssrn.com/abstract=4716427.
[120] Barczentewicz M. How the new interoperability mandate could violate the EU charter. Lawfare; 2023 [Internet][cited 2024 Apr 17]. Available from: https://www.lawfaremedia.org/article/how-the-new-interoperability-mandate-could-violate-the-eu-charter.
[121] Brown I, Korff D. Key points on DMA interoperability and encryption. Ian Brown. WordPress; 2022 [Internet][cited 2024 Apr 17]. Available from: https://www.ianbrown.tech/2022/04/01/key-points-on-dma-interoperability-and-encryption/.
[122] Morton FM, Crawford GS, Crémer J, Dinielli D, Fletcher A, Heidhues P, et al. Equitable interoperability: the "super tool" of digital platform governance. Yale J Regul 2023;40:1013.
[123] Anderlini J, Milani C. Emerging forms of sociotechnical organisation: the case of the fediverse. Digital platforms and algorithmic subjectivities. University of Westminster Press; 2022. p. 167–81. https://doi.org/10.16997/book54.m [Internet]Available from:.
[124] Newton C. Meta unspools threads. Platformer [Internet]. 2023 Jul 5; Available from: https://www.platformer.news/meta-unspools-threads/.
[125] Schauer F.F. Free speech: a philosophical enquiry. 1982.
[126] Greenawalt K. Free speech justifications. Columbia Law Rev 1989;89(1):119. Jan.
[127] Bluesky. Bluesky's moderation architecture. Bluesky; 2024 [Internet][cited 2024 Apr 2]. Available from: https://docs.bsky.app/blog/blueskys-moderation-architecture.